\documentclass[11pt]{article}
\usepackage{epsfig}
\usepackage{amsfonts,amssymb,amsmath}

\def \RTITLE {{\small R. Longo, K.-H. Rehren: How to remove the boundary}}
\def \BB {{\mathcal B}}
\def \HH {{\mathcal H}}
\def \RR {{\mathbb R}}      
\def \CC {{\mathbb C}}      
\def \ZZ {{\mathbb Z}}      
\def \Mb {\hbox{M\"ob}} 
\def \Hom {\mathrm{Hom}}
\def \VV {{\mathcal{V}}}
\def \Diff {{\mathrm {Diff}\;}}
\def \id {{\mathrm {id}}}
\def \inv {^{-1}}
\def \eps {\varepsilon}
\def \supp {\mathrm{supp}\;}
\renewcommand{\geq}{\geqslant}
\renewcommand{\leq}{\leqslant}
\def \QED {\hspace*{\fill}Q.E.D.}
\def \bea {\begin{eqnarray}}
\def \eea {\end{eqnarray}}

\title{\bf How to remove the boundary in CFT \\[2mm] \large -- an operator algebraic
  procedure} 

\author{Roberto Longo$^{1}$, \\[2mm]
Karl-Henning Rehren$^{2}$}

\begin{document}

\maketitle

\begin{center}
\scriptsize
$^{1}$ Dipartimento di Matematica, \\
Universit\`a di Roma ``Tor Vergata'', Via della Ricerca Scientifica 1,
00133 Roma, Italy \\[2mm]
$^2$ Institut f\"ur Theoretische Physik, Universit\"at G\"ottingen, \\
Friedrich-Hund-Platz 1, D-37077 G\"ottingen, Germany
\end{center}

\begin{abstract}
The relation between two-dimensional conformal quantum field theories 
with and without a timelike boundary is explored. 
\end{abstract}

\medskip

{\sl Dedicated to Klaus Fredenhagen on the occasion of his 60th birthday}

\normalsize

\section{Introduction}\label{sec1}

In \cite{LR2}, the authors have formulated boundary conformal field
theory in real time (Lorentzian signature) in the algebraic framework
of quantum field theory. BCFT is a local M\"obius covariant QFT $B_+$
on the two-dimensional Minkowski halfspace $M_+$ (given by $x>0$), which
contains a (given) local chiral subtheory $A$, e.g., the stress-energy
tensor. The reward of this approach was the surprisingly simple
formula ((\ref{relcomm}) below), expressing the von Neumann algebras
of local observables $B_+(O)$ in a double cone $O\subset M_+$ in terms
of an (in general nonlocal) chiral conformal net $B$ of localized
algebras associated with intervals along the boundary (the time axis
$x=0$). The net $B$ is M\"obius covariant and contains the local
chiral observables $A$:   
\bea
A(I)\subset B(I)
\eea
for each interval $I\subset\RR$. 

The reduction to a single chiral net is responsible for a kinematical
simplification, explaining, e.g., Cardy's observation \cite{C}
that in BCFT, bulk $n$-point correlation functions are linear
combinations of chiral $2n$-point conformal blocks.  

The algebra $B_+(O)$ is a relative commutant of $B(K)$ within $B(L)$, 
\bea\label{relcomm}
B_+(O) = B(K)'\cap B(L),
\eea
where $K\subset L$ are a pair of open intervals on the boundary $\RR$
such that the disconnected complement $L\setminus \overline K = I\cup
J$ is the set of advanced and retarded times $t\pm x$ associated
with points in $(t,x)\in O$ (see Fig.\ 1). Although the chiral net $B$
is not necessarily local, the intersections (\ref{relcomm}) do commute
with each other when two double cones are spacelike separated. 

The main result in \cite{LR2} is that every BCFT is contained in a
maximal (Haag dual) BCFT of the form (\ref{relcomm}).

\bigskip

\hskip40mm \epsfig{file=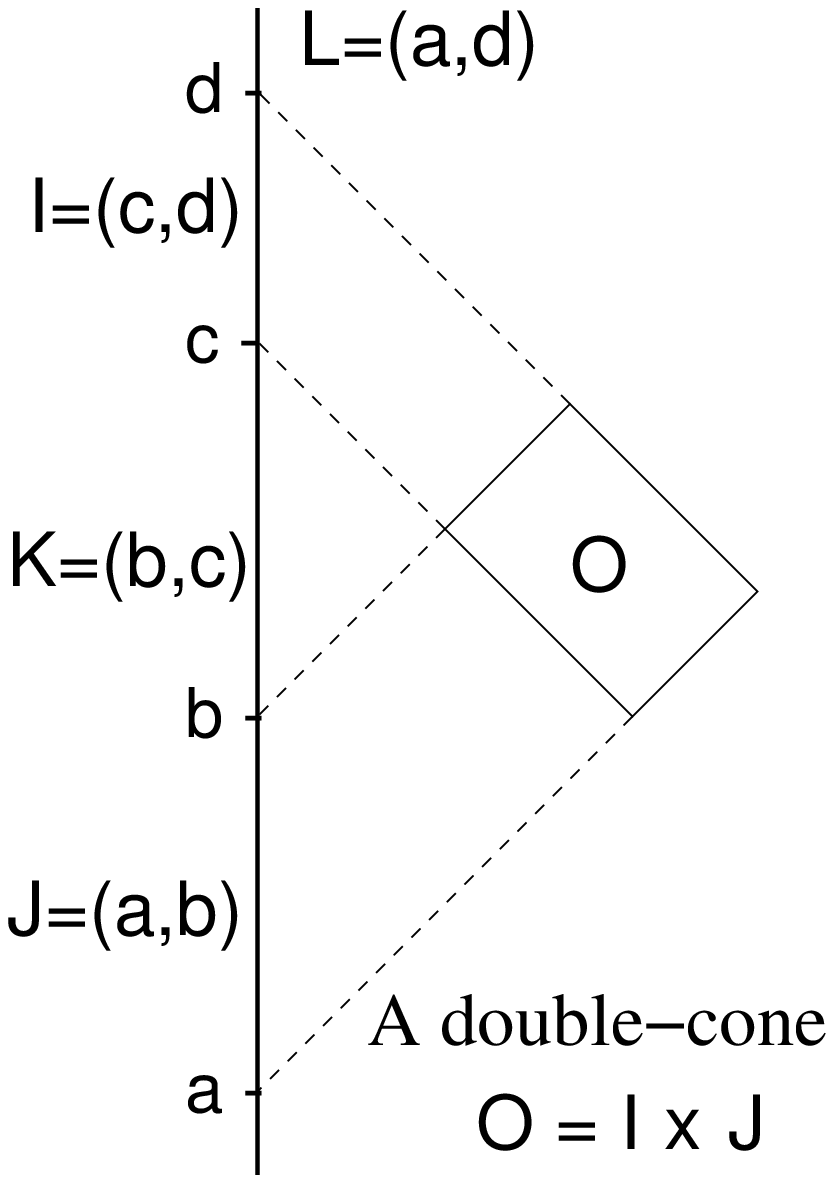,width=50mm} 
\nopagebreak

{\bf Figure 1:} Intervals on the boundary and double cones in the halfspace.

\bigskip

This leads to a somewhat paradoxical conclusion: on the one
hand, each local bulk observable is defined as a (special) observable
from a chiral CFT. Thus, superficially, the ``degrees of freedom'' of
a BCFT are not more than those of a chiral CFT, containing only a
single chiral component of the stress-energy tensor (Virasoro
algebra). One might argue that such a ``reduction of degrees of
freedom'' is a characteristic feature of QFT with a boundary. But this
point of view cannot be maintained, because on the other hand, the
resulting BCFT $B_+$ is locally equivalent to another CFT $B_{2D}$ on
the full two-dimensional (2D) Minkowski spacetime, which has all the
degrees of freedom of a 2D QFT, and in particular contains a full 2D
stress-energy tensor (two commuting copies of the Virasoro algebra). 
Even in the simplest case, when the chiral net $B$ on the boundary
coincides with $A$ (sometimes known as ``the Cardy case''), the
associated bulk QFT contains apart from the full 2D stress-energy 
tensor more (``non-chiral'') local fields that factorize into chiral
fields with braid group statistics. Locally, also the BCFT contains
the same fields. 
\medskip

This paradoxical situation is not a contradiction; it rather shows
that ``counting degrees of freedom'' of a QFT is an elusive
task. Trivially, there is no obstruction against a proper inclusion of
the form $\BB(\HH)\otimes\BB(\HH)\subset \BB(\HH)$ if $\HH$ is an
infinite-dimensional Hilbert space. But ``counting degrees of
freedoms'', e.g.\ by entropy arguments, requires the specification of
the Hamiltonian. The BCFT shares the Hamiltonian and ground state
(vacuum) of the chiral CFT, while the associated 2D CFT has a
different Hamiltonian and a different ground state. Thus, with respect
to different Hamiltonians, the spacetime dimension (measured through
some power law behaviour of the entropy) may assume different values
(1 or 2, in the present case). 

Looking at the issue from a different perspective, we may start from a
vacuum representation of the Virasoro algebra. The latter integrates
to a unitary projective representation of the diffeomorphism group of
the circle $\Diff(S^1)$, which contains the diffeomorphism group of an
interval $\Diff(I)$ as a subgroup. For two open intervals with
disjoint closures, there is a canonical identification between
$\Diff(I\cup J)$ and $\Diff(I)\times \Diff(J)$. In terms of the
stress-energy tensor $T$, this amounts to an isomorphism between $\exp
iT(f+g)$ and $\exp iT(f)\otimes \exp iT(g)$, when $f$ and $g$ have
disjoint support. It would be hard to see this local isomorphism
directly in terms of the Virasoro algebra.  

\medskip

The mathematical theorem underlying these facts is the well-known
Split Property \cite{DL}, which can be derived in local QFT in any
dimension under a suitable phase space assumption. In chiral local 
CFT, a sufficient assumption is the existence of the conformal
character $\mathrm {Tr}\; \exp-\beta L_0$. 

In the algebraic framework, the {\em chiral} observables of a BCFT
(e.g., the stress-energy tensor) localized in a double cone $O$ are
operators belonging to the von Neumann algebra $A_+(O) = A(I)\vee
A(J)$ where $I$ and $J$ are two open intervals of the time axis
(``advanced and retarded times'') such that $ t+x\in I$, $t-x\in J$
for $(t,x)\in O$ (this justifies the notation $O=I\times J$), and
$A(I)$ are the von Neumann algebras generated by the unitary
exponentials of chiral fields smeared within $I$. In contrast, the
{\em chiral} observables in a 2D CFT are operators in the algebra
$A_{2D}(O) = A_L(I)\otimes A_R(J)$ where $I$ and $J$ are regarded as
two open intervals of the lightcone axes, and $A_R(I)$ and $A_L(J)$
are  generated by left and right chiral fields. Our present
association between BCFT and 2D CFT applies to the case when $A_L(I) =
A_R(I)= A(I)$, i.e., the left chiral observables $A_L(I)\otimes 1$ are
isomorphic with the right chiral observables $1\otimes A_R(I)$, and
both are isomorphic with the chiral observables $A(I)$ of the BCFT.   

\medskip

Let $\HH_0$ denote the vacuum Hilbert space for the chiral CFT
described by the algebras $A(I)$. The split property states that if
$I$ and $J$ are two intervals with disjoint closures, there is a
canonical unitary $\VV:\HH_0\to\HH_0\otimes\HH_0$ implementing an
isomorphism 
\bea\label{splitiso}
\VV\big(A(I)\vee A(J)\big)\VV^* = A(I)\otimes A(J).
\eea

The split isomorphism does not preserve the vacuum vector, i.e.,
the canonical ``split vector'' $\Xi=\VV^*(\Omega\otimes\Omega)$ is an
excited state in $\HH_0$. By construction, the split state
$(\Xi,\cdot\Xi)$ on $A(I)\vee A(J)$ has the property that its
expectation values for either subalgebra $A(I)$ or $A(J)$ coincide
with those in the vacuum state, but the correlations between
observables $a_1\in A(I)$ and $a_2\in A(J)$ are suppressed: 
\bea\label{splitstate}
(\Xi\,,\,a_1a_2\,\Xi) = (\Xi\,,\,a_1\,\Xi)\;(\Xi\,,\,a_2\,\Xi) = (\Omega\,,\,a_1\,\Omega)\;(\Omega\,,\,a_2\,\Omega).
\eea
The split isomorphism depends on the pair of intervals $I$ and 
$J$. It trivially restricts to algebras associated with subintervals,
but it does not, in general, extend to larger intervals. When the
intervals touch or overlap, a split state and the split isomorphism
cease to exist.  

While the split isomorphism is well known, we discuss in this paper
its extension to ``non-chiral'' local observables, which do {\em not} 
belong to  $A(I)\vee A(J)$ in the BCFT, and to $A(I)\otimes A(J)$ in
the 2D CFT. 

\medskip

As a concrete demonstration for the resolution of the above ``paradox'', 
we present two simple but nontrivial examples where the algebraic
relations outlined can be easily translated into the field-theoretic
setting, i.e., we characterize the local algebras of the various QFTs
in terms of generating local Wightman fields. 

Let us translate (\ref{relcomm}) into the field-theoretic language. 
The intervals $I$ and $J$ shrink to the points $t\pm x$ when
$O=I\times J$ shrinks to a point $(t,x)$. Thus, we have to approximate
a field $\Phi(t,x)$ of the BCFT by observables in $A(L)$ (where the
interval $L$ approximates $(t-x,t+x)$ from the outside), that commute
with all fields localized in the interval $K$ (which approximates
$(t-x,t+x)$ from the inside). This will be done in Sect.\ 2. A crucial
point here is that generating the local algebra $A(L)$ involves
``non-pointwise'' operations, e.g., typical observables may be
exponentials of smeared field operators, so that an element of the
relative commutant is not necessarily localized in the disconnected
set $L\setminus \overline K = I \cup J$.   

\medskip

A second, somewhat puzzling feature of the algebraic treatment of BCFT
is the fact that the description of the local algebras $B_+(O)$ in
terms of the chiral boundary net (Eq.\ (\ref{relcomm})) is much
simpler than that of the local algebras $B_{2D}(O)$ of the associated
$2D$ conformal QFT without a boundary. The latter are (rather
clumsily) defined as Jones extensions of the tensor products
$A(I)\otimes A(J)$ in terms of a Q-system constructed from the
chiral extension $A\subset B$ with the help of $\alpha$-induction
\cite{KHR}.   

One purpose of this work is to present a more direct construction
of the 2D CFT without boundary from the BCFT. The obvious idea is to
take a limit as the boundary is ``shifted to infinity''. But we shall
do more, and establish the {\em covariant} local isomorphism between the
subnets $O\mapsto B_+(O)$ and $O\mapsto B_{2D}(O)$ as $O\subset O_0$,
i.e., the restriction of the AQFTs to any double cone $O_0$ within the
halfspace $x>0$, at finite distance from the boundary. 

\medskip

The main problem here is, of course, the enhancement of the conformal
symmetry, i.e., the reconstruction of the unitary positive-energy
representation of the two-dimensional conformal group $\Mb\times \Mb$
from that of the chiral conformal group $\Mb$. This is done by
a ``lift'' of the chiral M\"obius covariance of the local chiral net
$A$, using the split property which allows to ``embed'' the 2D chiral
algebra $A(I)\otimes A(J)$ into a local BCFT algebra $B_+(O)$. This
will be done in Sect.~3. The point is that only a single local
algebra of the BCFT is needed for this reconstruction of the 2D
conformal group and the full 2D CFT.

\medskip 

In Sect.~4, we show that the 2D CFT can also be obtained through a
limit where the boundary is ``shifted to the left'', or equivalently,
the BCFT observables are ``shifted to the right''. The translations in
the spatial direction ``away from the boundary'' do not belong to the
chiral M\"obius group of the BCFT. But they are at our disposal by the
previous lifting of the 2D M\"obius group into the BCFT. Therefore, we
can study the behavior of correlation functions in the limit of
``removing the boundary''. As we shift the boundary, the retarded and
advanced times are shifted apart from each other. The convergence of
the vacuum correlations of the BCFT to the vacuum correlations of the
2D CFT is therefore a consequence of the cluster behavior of vacuum
correlations of the chiral CFT $A$.  

\medskip

We add three appendices containing some related observations.

\section{Example}\label{sec2}
\setcounter{equation}{0}

The purpose of this section is to illustrate the construction
(\ref{relcomm}) in a field-theoretic setting. It is convenient to
assume the trivial chiral extension $B=A$ since even in this case
the construction (\ref{relcomm}) is nontrivial, i.e., non-chiral local
BCFT fields that factorize into nonlocal chiral fields can be
constructed from local chiral fields only. We exhibit local BCFT 
fields in a region $O=I\times J\subset M_+$ as ``neutral'' chiral
operators, that behave like products of ``charged'' chiral operators
localized in $I$ and $J$ in the limit of large distance from the
boundary. The limit of pointlike localization is also discussed, and
reproduces familiar vertex operators.

\medskip 

Consider the free $U(1)$ current $j$ with commutator
$[j(x),j(y)]=2\pi i\delta'(x-y)$ and charge operator $Q=(2\pi)\inv\int
j(x)dx$. The unitary Weyl operators $W(f)=e^{ij(f)}$ for real test
functions $f$ satisfy the Weyl relation  
\bea\label{weyl}
W(f)\,W(g) = e^{-i\pi\sigma(f,g)}\cdot W(f+g) = e^{-2\pi i\sigma(f,g)}
\cdot W(g)\,W(f)
\eea
and have the vacuum expectation value
\bea\label{ground}
\omega(W(f)) = e^{-i\pi\sigma(f_-,f_+)} = e^{-\frac
  12\int_{\RR_+}k\,dk \vert \hat f(k)\vert^2}
\eea
where the symplectic form is
\bea\label{symp}
\sigma(f,g)= \frac 12\int_\RR dx\;\big(f(x)\;g'(x) - f'(x)\;g(x)\big)
= \frac 1{2\pi i}\int_\RR k\,dk\;\hat f(-k)\;\hat g(k),
\eea
and $f_+$ ($f_-$) correspond to the restrictions to positive
(negative) values of $k$ of the Fourier transform $\hat f(k) =
\int_\RR dx\,e^{ikx}\,f(x)$. With these conventions, $W(f)\Omega$ is a 
state with charge density $-f'(x)$.   

The vacuum correlations of Weyl operators are
\bea\label{corr}
\omega(W(f_1)\cdots W(f_n)) =
e^{-i\pi\big(\sum_i\sigma(f_{i-},f_{i+}) + 2
  \sum_{i<j}\sigma(f_{i-},f_{j+})\big)}
\eea

The Weyl operators $W(f)$ with $\supp f\subset I$ define the local
von Neumann algebras of the chiral net $I\mapsto A(I)$. We fix a
double cone $O=I\times J\in M_+$. If $f$ is a test function that
vanishes outside $L$ and is constant in $K$, then $W(f)$ belongs to
$A(L)$ and commutes with $A(K)$ by (\ref{weyl}) and (\ref{symp}), hence
\bea\label{neut}
W(f)\subset B_+(O) = A(K)'\cap A(L). 
\eea
These are examples of operators that belong to $B_+(O)$ but (if
$f\vert_K\neq 0$) not to $A_+(O)= A(I)\vee A(J)$.

\medskip

Weyl operators can also be defined for smooth functions
$f$ such that $f'$ has compact support, and the relation
(\ref{weyl}) holds. Then $q= f(-\infty)-f(\infty)$ is called the
charge. However, $i\sigma(f_-,f_+)$ diverges, and the vacuum
expectation value (\ref{ground}) vanishes unless $q=0$ (see below).  
This implies that correlation functions (\ref{corr}) of charged Weyl
operators vanish whenever the total charge is non-zero (charge
conservation), while the IR divergences in each term in the exponent
of (\ref{corr}) cancel for neutral
correlations. The neutral Weyl operators (\ref{neut}) in $B_+(O)$ are
(up to a phase factor) products of charged Weyl operators with charge
densities localized in $J$ and in $I$. 

In the limit of sharp step functions $G_u(x)=q\cdot \theta(x-u)$
(requiring a regularization \cite{CRW}), the regularized Weyl
operators $W(G_u)$ become the well-known vertex operators of charge
$-q$ and scaling dimension $\frac 12q^2$ \cite{SSV}, which are
formally written as     
\bea 
V_{-q}(u) = {:\! \exp\Big(iq\int_u^{\infty} j(y)dy\Big) \!:}.
\eea  
Thus, as $O$ shrinks to a point $(t,x)\in M_+$, and $I$ and $J$ shrink
to the points $t+x$ and $t-x$, the (regularised) Weyl operators
$W(G_{t-x}-G_{t+x})$ behave as 
\bea
\Phi_{q}(t,x) = V_{q}(t+x) V_{-q}(t-x).
\eea
The correlation functions of vertex operators are given by 
\bea \langle\ldots\cdot V_{q_i}(u_i)\cdot\ldots\rangle =
\lim_{\eps\searrow 0}\;  \prod_{i<j}
\Big(\frac{-i}{u_i-u_j-i\eps}\Big)^{-q_iq_j} \eea
if $\sum_i q_i=0$, and $=0$ otherwise, from which the well-known anyonic
commutation relations can be read off. It is then easily seen
that $\Phi_{q_1}(t_1,x_1)$ commutes with $\Phi_{q_2}(t_2,x_2)$ 
when either $t_1+x_1 > t_2+x_2 > t_2-x_2 > t_1-x_1$ or when $t_2+x_2 >
t_1+x_1 > t_1-x_1 > t_2-x_2$, because in these cases the anyonic phase
factors cancel. It also commutes with $j(t_2\pm x_2)$ if $t_2\pm
x_2\neq t_1\pm x_1$. These are precisely the requirements for locality
of the fields $\Phi_q(t,x)$ among each other, and relative to the
conserved current 
\bea
j_0(t,x) = j(t+x) + j(t-x),\qquad j_1(t,x) = j(t+x) - j(t-x) 
\eea
defined for $x>0$, i.e., $\Phi_q$ and $j^\mu$ are local fields on the
halfspace $M_+$. The correlation functions of $n$ fields
$\Phi_{q_i}(t_i,x_i)$ are correlations of $2n$ vertex operators
($2n$-point conformal blocks).

\medskip 

After this digression to pointlike fields, let us resume the
study of the correlation functions (\ref{corr}) of the smooth
Weyl operators $W(f_i)\in B_+(O)$, and their behavior as $O$ is
shifted away from the boundary. We choose $n$ test functions of the 
form 
\bea\label{g-h}
f_i=G_i-H_i
\eea
where $G_i$, $H_i$ are smooth step functions with values $0$ at
$-\infty$ and $q_i$ at $+\infty$, such that $G_i'=g_i$ is supported in
$J$ and $H_i'=h_i$ is supported in $I$ (see Fig.\ 2).

\bigskip

\hskip10mm \epsfig{file=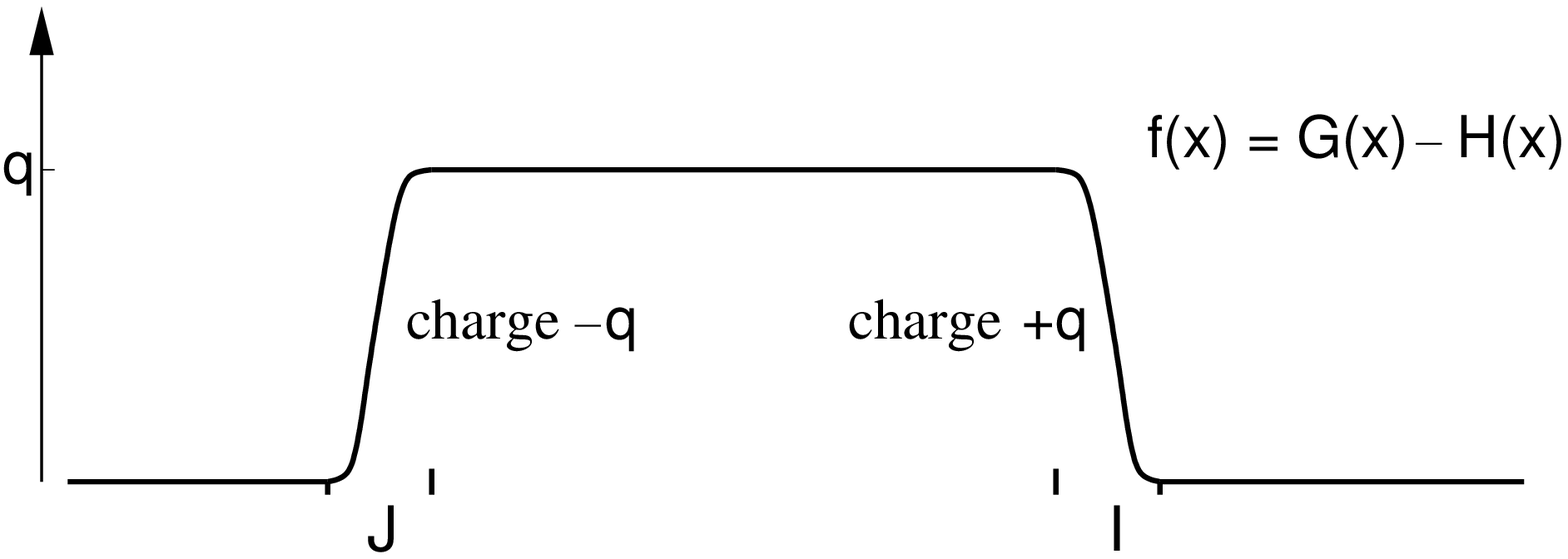,width=100mm} 
\nopagebreak

{\bf Figure 2:} A test function $f$ such that $W(f)$ belongs to
$B_+(O)$, but not to $A_+(O)$. $G$ and $H$ are smooth step functions,
$\supp G' \subset J$, $\supp H' \subset I$.

\bigskip

The neutral states $W(f_i)\Omega$ carry the charge $q_i$ in $I$ and the
charge $-q_i$ in $J$. 

The neutrality
condition for each Weyl operator $W(f_i)$ can be written 
\bea
\label{charge}
\int_\RR dx\;g_i(x) - \int_\RR dx\;h_i(x)= 0 \qquad\Leftrightarrow \qquad
\hat g_i(0) - \hat h_i(0) =0. 
\eea 

The exponent in (\ref{corr}) is a linear combination of terms of the form 
(using $\hat f_i = i(\hat g_i-\hat h_i)/k$) 
\bea
2\pi i\,\sigma(f_{i-},f_{j_+}) = \int_{\RR_+} \frac{dk}k
\int dx\,\big(g_i(x)-h_i(x)\big)\int dy\, \big(g_j(y)-h_j(y)\big)  
 \;e^{-ik(x-y)} \quad
\eea
which are IR finite because of (\ref{charge}). The separate
contributions from $g_i$ and $h_i$, however, are IR divergent. 
Therefore, we first regularize at $k=0$ by the subtraction
$e^{-ik(x-y)}\to e^{-ik(x-y)}-e^{-k/\mu}$ ($\mu>0$ arbitrary), 
which does not change the result because of (\ref{charge}), and then
compute the contributions from $g$ and $h$ separately. 

We are interested in the behavior of the correlation function
(\ref{corr}) as $O$ is shifted away from the boundary. This means that
the functions $g_i$ are shifted by a distance $a$ to the left, and
$h_i$ are shifted by the same distance to the right. The $g$-$g$
contributions and the $h$-$h$-contributions to
$\sigma(f_{i-},f_{j_+})$ are obviously invariant under this shift,
while in the mixed $h$-$g$ contributions $x-y$ is replaced by $x-y+2a$:
\bea
2\pi i\,\sigma_{h_i,g_j}(a) := - \int_I dx\, h_i(x) \int_J dy\, g_j(y)  
\int_{\RR_+} \frac{dk}k \Big[e^{-ik(x-y+2a)}-e^{-k/\mu}\Big] 
\eea
and similar for the $g$-$h$ contributions. The last integrand
can be split into two parts:
\bea
\big(e^{-ik(x-y+2a)}-1\big)e^{-k/\mu} \;+\;
e^{-ik(x-y+2a)}\big(1-e^{-k/\mu}\big) 
\eea
so that the first contribution to the momentum integral equals 
\bea
-\log\big(1+i\mu(x-y+2a)\big)
\eea
while the second (distributional) contribution is of order $O(a\inv)$
in the limit of large $a$. Because the remaining integrals have
compact support, we obtain    
\bea
\lim_{a\to\infty} \sigma_{h_i,g_j}(a) = q_iq_j\cdot \log(2i\,a\mu)
+ O(a\inv).
\eea
Together with the $g$-$h$ contributions $q_iq_j\cdot \log(-2i\,a\mu)$,
these terms in the exponent of (\ref{corr}) cumulate up to
the factor   
\bea
\prod_{i} (2a\mu)^{-q_i^2} \prod_{i<j} (2a\mu)^{-2q_iq_j} = (2a\mu)^{-q^2},
\eea
where $q=\sum_i q_i$ is the total charge within $I$. Thus (\ref{corr})
vanishes in the limit $a\to\infty$ if $q\neq 0$, enforcing ``chiral
charge conservation'' in the limit. If $q=0$, the remaining $g$-$g$
and $h$-$h$ contributions yield
\bea
\lim_{a\to\infty} \omega\big(W(f_1)\cdots W(f_n)\big) = 
\omega\big(W(G_1)\cdots W(G_n)\big)\cdot \omega\big(W(-H_1)\cdots
W(-H_n)\big) \qquad 
\eea
involving charged Weyl operators. These expressions are well-defined
(and independent of $\mu$) because $\sum_i G_i$ and $\sum_i H_i$ are
neutral precisely due to $q=0$.   

\medskip

The factorization of the vacuum correlations in the limit $a\to\infty$
is the desired feature we wanted to illustrate by this example. In the
limit, $W(f_i)$ have the same correlations as $W(-H_i)\otimes W(G_i)$,
which are charged observables of the associated 2D CFT. 
Notice that in the limit of sharp test functions (see above), one
obtains  
\bea
V_{q}(t+x) \otimes V_{-q}(t-x)
\eea
which are local fields in the entire two-dimensional Minkowski spacetime $M^2$.

\medskip

{\bf Remark:} The above construction can be generalized to the $SU(2)$
current algebra. The Frenkel-Kac representation of $SU(2)$ currents at
level 1 is given by $j^3\equiv j$ and $j^\pm(x) = j^1(x) \pm i j^2(x) =
V_{\pm \sqrt2}(x)$. Then $V_{q}(x)\cdot V_{-q}(y)$ commutes with
$V_{q'}(w)$ at $w\neq x,y$ provided $qq'\in \ZZ$. Hence the field 
\bea
\Phi_{\frac 12\sqrt2}(t,x) = V_{\frac 12\sqrt2}(t+x)\cdot 
V_{-\frac 12\sqrt2}(t-x)
\eea
is local (as before) and relatively local w.r.t.\ the conserved currents 
$j^a$ ($a=1,2,3$)
\bea
j^a_0(t,x) = j^a(t+x) + j^a(t-x),\quad j^a_1(t,x) = j^a(t+x) -
j^a(t-x).\qquad
\eea
$\Phi_{\frac 12\sqrt2}(t,x)$ is a neutral combination of charged
primary fields of dimension $\frac 14$, transforming in the
spin-$\frac12$ representation of $SU(2)$, localized at $t+x$ and
$t-x$. The description of this model in terms of
smooth Weyl operators is rather straightforward, see e.g., \cite{BMT}:
Weyl operators with integer multiples of the charge $\sqrt 2$ belong to
$A(I)$, while operators with half-integer multiples of the charge
$\sqrt 2$ in $I$ and in $J$ belong to $A(K)'\cap A(L)$. 

\medskip

The mechanism of ``charge separation'' described here for obtaining 
elements of $B_+(O)$ that do not belong to $A_+(O)$ is very general
\cite{LR2}, although in general it cannot be formulated in terms of
Weyl operators. In Sect.\ 4 we shall show that also the factorization
behavior far away from the boundary is a general feature, which allows
to recover the 2D CFT from the BCFT. 

\section{Reconstruction of the 2D symmetry}\label{sec3}
\setcounter{equation}{0}

We work in this section with a fixed ``chiral extension'' $A\subset B$. 
Here, $A$ is a Haag dual M\"obius covariant local net $\RR\supset
I\mapsto A(I)$ of von Neumann algebras on its vacuum Hilbert space
$\HH_0$, satisfying the split property and having finitely many
irreducible DHR sectors of finite dimension (these properties together
are called ``complete rationality'' \cite{KLM}; in the case of
diffeomorphism covariant nets, Haag duality = strong additivity is a
consequence of the other properties \cite{LX}. The fact that the $U(1)$
Weyl algebra in Sect.~2 is {\em not} completely rational, indicates
that the results to be reported in this section hold also in more
general situations).    

$B$ is a M\"obius covariant net $\RR\supset I\mapsto B(I)$ on its
vacuum Hilbert space $\HH_0^B$ such that for each $I$ the inclusion
$A(I)\subset B(I)$ holds and is an irreducible subfactor, which has
automatically finite Jones index \cite{KL} equal to the statistical
dimension of the (reducible) representation of $A$ on $\HH_0^B$
\cite{LR1}. The net $B$ may be non-local, but is required to be
relatively local w.r.t.\ $A$.  

If only $A$ is specified, the irreducible chiral extensions $B$ of $A$
can be classified in terms of Q-systems of $A$ \cite{LR1}. The
complete classification has been computed for $A$ the Virasoro nets
with central charge $c<1$ (and implicitly also for the $SU(2)$ current
algebras) in \cite{KLPR}. 

\medskip

With $A\subset B$ one can associate a boundary CFT $B_+$ on the
halfspace $M_+$ and a two-dimensional CFT $B_{2D}$ on Minkowski spacetime
$M^2$. To describe the former, we introduce a convenient notation (see
Fig.\ 1). 
For any quadruples of four real numbers such $a<b<c<d$ we define 
$I=(c,d)$, $J=(a,b)$, $K=(b,c)$, $L=(a,d)$, and $O =
\{(t,x):\; t+x\in I,\, t-x \in J\}\subset M_+$. Every double cone
$O\subset M_+$ is of this form and determines $I,J,K,L$, and similarly
every pair of open intervals $J<I$ (``$I$ is to the right = future of
$J$'') determines $K,L$, and $O=I\times J$. 

Then the BCFT associated with $A\subset B$ is the net (\ref{relcomm}),
i.e., $O\mapsto B_+(O)= B(K)'\cap B(L)$. We have shown in \cite{LR2}
that $B_+(O)$ contains $A_+(O) = A(I)\vee A(J)$ as a subfactor with
finite index, $B_+$ is local and Haag dual on $M_+$, every Haag dual
BCFT with chiral observables $A$ arises in this way (namely the chiral
extension $B$ can be recovered from the BCFT), and every non-Haag-dual
local BCFT net is intermediate between $A_+$ and $B_+$. If $B=A$,
$B_+(O)$ equals the four-interval subfactor $A(E)\subset A(E')'$ on
the circle \cite{KLM} ($E=I\cup J$).     

The 2D CFT $B_{2D}$ associated with $A\subset B$ has been constructed
in \cite{KHR}. Its local algebras are extensions (with finite Jones
index) of the tensor products $A(I)\otimes A(J)$, specified in terms
of a Q-system constructed from the chiral extension $A\subset B$
with the help of $\alpha$-induction. 

We know from \cite{LR2} that $B_+$ and $B_{2D}$ are locally
isomorphic, i.e., for each $O\subset M_+$ there is an isomorphism
$\varphi^O: B_+(O)\to B_{2D}(O)$ such that
\bea\label{lociso}
 \varphi^O\big(B_+(O_1)\big) = B_{2D}(O_1)\quad \hbox{for all} \;
O_1\subset O. 
\eea 
However, the Hilbert space and the vacuum state for the two theories
are very different. 

In this section, we wish to understand the relation between these two
nets, by giving an alternative construction of the 2D CFT directly from 
the BCFT. The crucial point is the construction of the enhanced
M\"obius symmetry of the 2D CFT, and its ground state (the 2D vacuum)
which is different from the BCFT vacuum.

\medskip

We first construct the Hilbert space $\HH_{2D}$ for the 2D CFT. We
choose a fixed reference double cone $O_0=I_0\times J_0\subset M_+$. 
The subfactor $A_+(O_0)=A(I_0)\vee A(J_0) \subset B_+(O_0)=B(K_0)'\vee
B(L_0)$ is irreducible with finite index \cite{LR2}, and hence has a unique
conditional expectation $\mu:B_+(O_0)\to A_+(O_0)$, which is automatically
normal and faithful. Let $\Xi\in\HH_0$ be the canonical split vector
for $A(I_0)\vee A(J_0)$ as in (\ref{splitstate}). The split state
$\xi=(\Xi,\cdot\Xi)$ on $A_+(O_0)$ extends to the state
$\hat\xi=\xi\circ\mu$ on $B_+(O_0)$. Let $\hat\HH$, $\hat\Xi$ and
$\hat\pi$ denote the GNS Hilbert space, GNS vector and GNS
representation for $(B_+(O_0),\hat\xi)$. We also write $\vert b\rangle$
for $\hat\pi(b)\hat\Xi$. Let us analyze the structure of $\hat\HH$. 

\medskip

The structure of $B_+(O_0)$ has been described in \cite{LR2}. 
By complete rationality, $A$ has finitely many irreducible
superselection sectors \cite{KLM}. Choose for each irreducible sector 
of $A$ a representative DHR endomorphism \cite{DHR} $\sigma$ localized
in $I_0$, and a representative $\tau$ localized in $J_0$. (For the vacuum
sector, $\sigma=\tau=\id$. $\bar\sigma$ and $\bar\tau$ are the
representatives of the conjugate sector.) Then the elements of
$B_+(O_0)$ are (weak limits of) sums of operators of the form $\iota(a_1
a_2) \cdot \psi$ where $\iota$ is the injection $A\to B$, $a_1\in
A(I_0)$, $a_2\in A(J_0)$, and $\psi \in B(L_0)$ generalize the Weyl
operators $W(f)$ (\ref{g-h}) of Sect.~2: they are (for each pair
$\sigma,\tau$) ``charged'' intertwiners in
$\Hom(\iota,\iota\sigma\bar\tau)\cap B(K_0)'$. We may express these
intersections in a different way: Let $\alpha_\rho^\pm$ denote the
endomorphisms of $B$ extending the DHR endomorphisms $\rho$ of $A$ 
by ``$\alpha$-induction'' \cite{LR1}, where $\alpha^+_\rho$
($\alpha^-_\rho$) acts trivially on $b\in B$ localized to the right =
future (left = past) of the interval where $\rho$ is localized. Thus
$\alpha_{\sigma_1}^-\alpha_{\bar\tau_1}^+$ acts trivially on $B(K_0)$, 
because $J_0<K_0<I_0$. Hence 
\bea\label{chint}
\Hom(\iota,\iota\sigma\bar\tau)\cap B(K_0)' =
\Hom(\id_B,\alpha_{\sigma}^-\alpha_{\bar\tau}^+).
\eea
(For an alternative characterization of the charged intertwiners
by means of an eigenvalue condition, see App.~\ref{appeig}.) 
If $O_1\subset M_+$ is another double cone in the halfspace, the
algebra $B_+(O_1)$ is generated by
$A(I_1)\vee A(J_1)$ and charged intertwiners 
\bea\label{chint1}
\psi_1 = \iota(u\times \bar u)\cdot\psi \in \Hom(\id_B,\alpha_{\sigma_1}^-\alpha_{\bar\tau_1}^+)
\eea
with
unitary charge transporters $u\in\Hom(\sigma,\sigma_1)$ and $\bar
u\in\Hom(\bar\tau,\bar\tau_1)$, where $\sigma_1$ is localized in $I_1$
and $\bar\tau_1$ is localized in $J_1$.

E.g., if $B=A$ (the ``Cardy case''), the charged intertwiners 
(generalizing the Weyl operators $W(f)$ in (\ref{g-h}) of Sect.~2) are of
the form $\psi\in \Hom(\id,\sigma\bar\tau)$. This implies that $\tau$
and $\sigma$ are representatives of the same sector. 
Thus, the charges of BCFT fields are in 1:1 correspondence
with the DHR sectors of $A$. 

In the general case, when $\psi$ and $\psi'$ are two charged
intertwiners, $\mu(\psi'\psi^*)$ is an intertwiner  
$\in \Hom(\sigma'\bar\tau',\sigma\bar\tau)\cap (A(I_0)\vee A(J_0))$. This
space is zero unless $\sigma'=\sigma$ and $\tau'=\tau$, and
$\Hom(\sigma\bar\tau,\sigma\bar\tau) \cap (A(I_0)\vee A(J_0)) = \CC\cdot 1$
\cite{L}. 
Therefore, we may choose (for each pair $\sigma,\tau$) a basis of
charged intertwiners $\psi$  which is orthonormal w.r.t.\ the inner
product $\mu(\psi'\psi^*)$.  

\medskip 

{\bf Lemma 1:} {\sl The subspaces $\hat\HH_\psi$ of $\hat\HH$ spanned by
$\vert\psi^*\cdot \iota(A(I_0) \vee A(J_0))\rangle$ are mutually
orthogonal. Each subspace $\hat\HH_\psi$ factorizes as a
representation of $A_+(O_0)$ according to 
\bea\label{hpsi}
 \hat\HH_\psi \cong \HH_\sigma\otimes \HH_{\bar\tau} 
\eea
where $\HH_\sigma$ and $\HH_{\bar\tau}$ carry the representations
$\sigma$ and $\bar\tau$ of $A(I_0)$ and $A(J_0)$, respectively. 
}

\medskip

{\it Proof:} The computation of matrix elements in a dense
set of vectors
\bea
\langle \psi^*\cdot\iota(a''_1a''_2)\vert \,\hat\pi\big(\iota(a_1a_2)\big)\,
\vert \psi^*\cdot\iota(a_1'a_2')\rangle =
\big(\Xi\,,\,a''_1{}^*a''_2{}^*\,\mu\big(\psi\,\iota(a_1a_2)\,\psi^*\big)\,a_1'a_2'\,\Xi\big) =
\qquad \\
= \big(\Xi\,,\,a''_1{}^*a''_2{}^*\,\sigma\bar\tau(a_1a_2)\,a_1'a_2'\,\Xi\big) =
\big(a''_1\,\Omega\,,\,\sigma(a_1)\,a_1'\,\Omega\big)\cdot
\big(a''_2\,\Omega\,,\,\bar\tau(a_2)\,a_2'\,\Omega\big)
\nonumber \eea
proves the claim. \QED

\medskip

We may therefore identify the vectors $\vert\psi^*
\iota(a_1'a_2')\rangle$ with $a_1'\Omega\otimes
a_2'\Omega\in\HH_{\sigma}\otimes\HH_{\bar\tau}$ in 
the representation $\sigma\otimes\bar\tau$ under the split
isomorphism, such that in particular, the GNS vector
$\hat\Xi=\vert 1\rangle\in\hat\HH$ corresponds to the 2D vacuum vector
$\Omega\otimes\Omega\subset \HH_0\otimes\HH_0$. We write
the extended Hilbert space $\hat\HH$ in the form    
\bea\label{h2d} 
\hat\HH \equiv \HH_{2D} \cong \bigoplus\nolimits_{\sigma,\tau} Z_{\sigma,\tau} \;
\HH_\sigma\otimes \HH_{\bar\tau} 
\eea 
(the ``2D Hilbert space''). The nonnegative integer multiplicities are
\bea\label{dim}
Z_{\sigma,\tau} = \dim \Hom(\alpha_{\tau}^+,\alpha_{\sigma}^-)
\eea
by the above characterization (\ref{chint}) of the spaces of charged
intertwiners. 
The chiral factorization (\ref{h2d}) of the GNS construction from the
extended state $\xi\circ\mu$ may be viewed as the remnant of the original
``splitting behavior'' of the split vector $\Xi$. 

As shown in \cite{LR2} by comparison of the Q-system, the local
subfactor $\hat\pi(A_+(O_0)) \subset \hat\pi(B_+(O_0))$ on $\hat\HH$
is isomorphic to $A(I_0)\otimes A(J_0)\subset B_{2D}(O_0)$ constructed
in \cite{KHR}. We may therefore consistently denote also the former by
$A_{2D}(O_0)\subset B_{2D}(O_0)$. 

\medskip

Next, we construct the action of the 2D M\"obius group on $\HH_{2D}$,
by a ``lift'' of the M\"obius transformations of the chiral net $A$,
using the split isomorphism and the conditional expectation $\mu$. 
The action of $\Mb\times \Mb$ on $\HH_{2D}$ will then be used to
define $B_{2D}(O)$ as the images of the reference algebra
$B_{2D}(O_0)$ under a 2D M\"obius transformation $g=(g_1,g_2)$ taking
$O_0$ to $O$. 

The 2D M\"obius group $\Mb\times \Mb$ is unitarily represented in the
vacuum Hilbert space $\HH_0$ of the chiral net $A$ by $U_+U_-$, the
preimage of $U_0\otimes U_0$ on $\HH_0\otimes\HH_0$ under the split
isomorphism. (See App.~\ref{appmod}, how $U_+$ and $U_-$ can be obtained by
modular theory directly on the boundary Hilbert space.) We need to
lift $U_+U_-$ to $\HH_{2D}$.

Let $\Sigma_I\subset \Mb$ denote the connected semigroup taking the
interval $I$ into itself, generated by the one-parameter
subgroup preserving $I$ and two one-parameter semigroups fixing either
of its endpoints. Then $\Sigma = \Sigma_{I_0}\times\Sigma_{J_0}\subset
\Mb\times\Mb$ is the connected semigroup taking the reference double
cone $O_0$ into itself.  

For $g=(g_1,g_2)\in\Sigma$, the adjoint action of $U_+(g_1)U_-(g_2)$
on $a_1\in A(I_0)$, $a_2\in A(J_0)$ is given by the {\em independent} 
(= product) action of the chiral M\"obius transformations given by
geometric automorphisms $\alpha_g$ of the chiral net $A$:
\bea\label{alpha}
\alpha^+_{g_1}\alpha^-_{g_2}(a_1\cdot a_2) =
\alpha_{g_1}(a_1)\cdot\alpha_{g_2}(a_2),
\eea
We extend these endomorphisms of $A_+(O_0)$ to
endomorphisms of $B_+(O_0)$ by
\bea\label{beta} 
\beta^+_{g_1}\beta^-_{g_2}\big(\iota(a_1a_2)\cdot\psi\big) :=
\iota\big(\alpha_{g_1}(a_1)\alpha_{g_2}(a_2)
\big)\cdot\iota\big(z^\sigma(g_1)z^{\bar\tau}(g_2)\big)\cdot \psi.
\eea
Here $z^\rho(g)\in \Hom(\rho,\alpha_g\rho\alpha_g\inv)$ are the
unitary cocycles \cite{FRS2,L} $z^\rho(g) = U_0(g)U_\rho(g)^*\in A$
where $U_0$ and $U_\rho$ are the representations of the M\"obius group
in the vacuum representation and in the DHR representation $\rho$. 

\medskip

{\bf Proposition 1:} {\sl (i) The maps $\beta^+_{g_1}\beta^-_{g_2}$
  defined by (\ref{beta}) for $g\in\Sigma$ are homomorphisms from
  $B_+(O_0)$ onto $B_+(g\,O_0)\subset B_+(O_0)$. \\[1mm] 
  (ii) For $O_1\subset O_0$ we have
  $\beta^+_{g_1}\beta^-_{g_2}\big(B_+(O_1)\big) = B_+(g\,O_1)$, i.e.,
  $\beta^+_{g_1}\beta^-_{g_2}$ ``act geometrically inside
  $B_+(O_0)$''. \\[1mm] 
  (iii) $\beta^+_{g_1}\beta^-_{g_2}$ respect the group composition law
  within the semigroup $\Sigma$.   \\[1mm] 
  (iv) The conditional expectation $\mu$ intertwines
  $\beta^+_{g_1}\beta^-_{g_2}$ with $\alpha^+_{g_1}\alpha^-_{g_2}$. }

\medskip

{\it Proof:} (i) The homomorphism property follows from the
composition and conjugation laws of charged intertwiners \cite{LR2}
and the intertwining and localization properties of the operators and
endomorphisms involved. The statement about the range is just a
special case of (ii). \\  
(ii) It is sufficient to show that a charged intertwiner 
$\psi_1\in B_+(O_1)$ is mapped to a charged intertwiner in $B_+(g\,O_1)$. By
virtue of (\ref{chint1}), we compute 
\bea \beta^+_{g_1}\beta^-_{g_2}(\psi_1) =
\iota\big(\alpha_{g_1}(u)z^\sigma(g_1)\alpha_{g_2}(\bar
u)z^{\bar\tau}(g_2)\big)\cdot \psi.
\eea
Then the claim follows, because $\alpha_{g_1}(u)\,z^\sigma(g)\in
\Hom(\sigma,\alpha_{g_1}\sigma_1\alpha_{g_1}\inv)$, and 
$\alpha_{g_1}\sigma_1\alpha_{g_1}\inv$ is localized in $g_1\, I_1$, and
similarly $\alpha_{g_2}\bar\tau_1\alpha_{g_2}\inv$ is localized in
$g_2\, J_1$.\\  
(iii) The group composition law follows from the cocycle
properties \cite{FRS2,L} of $z^\rho$. \\ 
(iv) The intertwining property of $\mu$ is due to the fact that $\mu$ 
annihilates all charged intertwiners except the neutral one
($\sigma=\bar\tau=\id$).\QED 

\medskip 

Next, we adapt a well-known Lemma about the implementation of
(groups of) automorphisms to the case of (semigroups of) endomorphisms.

\medskip 

{\bf Lemma 2:} {\sl  Let $M$ be a von Neumann algebra on a Hilbert
  space $\HH$ with a cyclic and separating vector $\Psi$. Let $\beta$
  be an endomorphism of $M$, preserving the state $(\Psi,\cdot\Psi)$. 
  Then the closure of the map $m\Psi\mapsto \beta(m)\Psi$ is an
  isometry $U_\beta$. If $\Psi$ is cyclic also for $\beta(M)$, then
  $U_\beta$ is unitary. For two endomorphisms $\beta$, $\beta'$ with
  the same properties, such that $\Psi$ is cyclic for $\beta(M)$, one
  has $U_{\beta'\beta} = U_{\beta'}U_{\beta}$. }  

\medskip

{\it Proof:} That $U_\beta$ is an isometry is an obvious consequence
of the invariance of the state. Since $\beta(M)\Psi$ is a dense
subset, the $U_\beta$ is surjective, hence unitary. For the last
statement it is sufficient to notice that $U_{\beta'}$ is densely
defined on $\beta(M)\Psi$. 
\QED 

\medskip 

We apply the Lemma to the endomorphisms $\beta^+_{g_1}\beta^-_{g_2}$
of $B_+(O_0)$. Using (iv) of Prop.~1, we see that
$\beta^+_{g_1}\beta^-_{g_2}$ leave the GNS state $(\hat\Xi,\cdot\hat\Xi)$
invariant because the split state $(\Xi,\cdot\Xi)$ on $A_+(O_0)$ is
invariant under $\alpha^+_{g_1}\alpha^-_{g_2}$. The vector $\hat\Xi$
is cyclic and separating for each $\hat\pi\big(B_+(O_1)\big)$
($O_1\subset O_0$) because $\mu$ is faithful and $\Xi$ is cyclic and
separating for each $A_+(O)$, which in turn follows by the split
isomorphism because $\Omega$ is cyclic and separating for $A(I_1)$ and
for $A(J_1)$. Thus, Lemma~2 applies:

\medskip

{\bf Corollary 1:} {\sl The homomorphisms 
$\beta^+_{g_1}\beta^-_{g_2}$ induce unitary
operators on $\hat\HH=\HH_{2D}$, which satisfy the group composition
law within the semigroup $\Sigma$. Together with the inverse unitary
operators, they generate a covering representation $\hat U(g_1,g_2) =
\hat U_+(g_1)\hat U_-(g_2)$ of $\Mb\times \Mb$ on $\HH_{2D}$.}

\medskip  

The last statement is due to the fact that $\Sigma$ and its inverse
generate $\Mb\times\Mb$, and the group law within $\Sigma$ secures the
commutation relations of the Lie algebra. 

By construction, for $g=(g_1,g_2)\in\Sigma$, $\hat U(g_1,g_2)$ on the
subspace $\HH_\psi$ is equivalent to $U_\sigma(g_1)\otimes U_{\bar\tau}(g_2)$ 
on $\HH_\sigma\otimes\HH_{\bar\tau}$ under the isomorphism (\ref{hpsi}). 
By (ii) of Prop.~1, the adjoint action of $\hat U(g_1,g_2)$ 
takes $B_+(O_1)$ to $B_+(g\,O_1)$ for $O_1\subset O_0$.

By constructing $U_+U_-$, we have thus furnished the local subnet
$O_0\supset O_1\mapsto B_+(O_1)$ of the BCFT with a covariant
``two-dimensional re-interpretation''. In the representation $\hat\pi$
on $\hat\HH=\HH_{2D}$, this is precisely the local isomorphism
$\varphi^{O_0}$ referred to in (\ref{lociso}). The present discussion
shows that $\varphi^{O_0}$ intertwines the global 2D M\"obius
covariance with a ``hidden'' symmetry of the BCFT, which is induced by the
extended split state $\hat\xi$ and acts locally geometric. 

\medskip

We now define for arbitrary double cones $O\subset M^2$ the associated
local algebras of the 2D conformal net on $\HH_{2D}$ by varying
$g=(g_1,g_2)\in \Mb\times\Mb$ in the connected neighborhood of unity
for which $g\,O_0\subset M^2$, and putting
\bea\label{b2d}
B_{2D}(O) := \hat U(g_1,g_2)\; 
B_{2D}(O_0)\; \hat U(g_1,g_2)^* \quad
\hbox{if}\quad O=g\, O_0\subset M^2.\;
\eea
For $O\subset O_0$, this coincides with
$\hat\pi\big(\beta^+_{g_1}\beta^-_{g_2}(B_+(O_0))\big) 
= \hat\pi\big(B_+(O)\big)$ by virtue of (ii) of Prop.~1. 
Notice that $B_{2D}(g\,O_0)$ is uniquely defined as long as 
$O= g\,O_0\subset M^2$ because in this case any two $g$ with the same
image $g\,O_0$ differ by an element of $\Sigma$, while it requires the
passage to a covering space when $M^2$ is conformally completed. 

\medskip 

{\bf Theorem 1:} {\sl The net of von Neumann algebras $O\mapsto
  B_{2D}(O)$ defined by (\ref{b2d}) is covariant, isotonous, and
  local.}

\medskip

{\it Proof:}  The covariance is by construction. Isotony and
locality of the 2D net follow from the geometric action inside $O_0$,
(ii) of Prop.~1, and the fact that every pair of double cones in $M^2$
such that either $O_1\subset O_2$ are $O_1\subset O_2'$ can be moved
inside $O_0$ by a M\"obius transformation, where we know (from the
boundary CFT) that isotony and locality hold. \QED

\medskip

{\bf Corollary 2:} {\sl The extension $A_{2D}\subset B_{2D}$ is
  isomorphic to the extension constructed in \cite{KHR}. } 

\medskip

{\it Proof:}  Since the local subfactor $A_{2D}(O_0)\subset B_{2D}(O_0)$
constructed in \cite{KHR} is isomorphic to $A_+(O_0)\subset B_+(O_0)$,
and the isomorphism intertwines the representations of the 2D M\"obius
group, the global isomorphism follows. \QED

\medskip

We have associated with the BCFT a 2D local CFT,
that is locally isomorphic. The association is intrinsic in the sense
that it requires only the subnet $O_0\supset O_1\mapsto B_+(O_1)$
together with the covariance of the DHR sectors of the underlying
chiral CFT $A$. 

It should be noticed that the construction is up to unitary
equivalence independent of the choice of the reference double cone
$O_0\subset M_+$. The reason is essentially that the charge structure
of $B(K)'\cap B(L)$ exhibited by the multiplicities $Z_{\sigma,\tau}$
in (\ref{h2d}) is independent of the pair $K\subset L$. 

\medskip

We conclude this section with an observation concerning diffeomorphism
covariance: 

\medskip

{\bf Proposition 2:} {\sl If $A\subset B$ is a chiral extension of a
  diffeomorphism covariant chiral net $A$, then the (possibly non-local) chiral net $B$, the BCFT
  net $B_+$ defined by (\ref{relcomm}), and the 2D net $B_{2D}$
  associated with $B_+$ by Thm.~1 are also diffeomorphism covariant.}

\medskip

{\it Proof:} The chiral net $A$ is diffeomorphism covariant if for a
diffeomorphism $\gamma$ of $S^1$ there is a unitary operator
$w_\gamma$ on $\HH_0$ such that $u_\gamma A(I) u_\gamma^* =
A(\gamma\,I)$. Haag duality of $A$ implies that if $\gamma$ is
localized in an interval $I$ (i.e., acts trivially on the complement),
then $w_\gamma$ is an observable in $A(I)$. 

For a chiral extension $A\subset B$ we claim that if $\gamma$
is localized in $I_0$, then for $I_1\subset I_0$ one has
$\iota(w_\gamma) B(I_1)\iota(w_\gamma^*) = B(\gamma I_1)$, i.e.,
$\iota(w_\gamma)$ implement the local diffeomorphisms. Namely,
$B(I_1)$ is generated by 
$\iota\big(A(I_1)\big)$  and $v_1=\iota(u)\cdot v$ where $v\in B(I_0)$  
is the canonical charged intertwiner $v\in\Hom(\iota,\iota\theta)$ 
for the canonical DHR endomorphism $\theta$ localized in $I_0$ 
\cite{LR1} (see also App.~\ref{appeig}), and $\theta_1$ is an
equivalent DHR endomorphism localized 
in $I_1$. We find 
\bea
\iota(w_\gamma)\;v_1\;\iota(w_\gamma^*) = \iota(w_\gamma u
\theta(w_\gamma^*))\cdot v.
\eea
Now, $w_\gamma u \theta(w_\gamma^*)\in
\Hom(\theta,\gamma\theta_1\gamma\inv)$, and $\gamma\theta_1\gamma\inv$
is localized in $\gamma\,I_1$. This proves the claim. The
diffeomorphism covariance of the chiral net $B$ follows because 
the diffeomorphisms localized in $I_0$ together with the M\"obius group
generate the diffeomorphism group of $S^1$. 

The argument for the boundary CFT and for the 2D CFT are very
similar: we first show that for diffeomorphisms
$\gamma=\gamma_1\gamma_2$ where $\gamma_1$ is localized in $I_0$ and
$\gamma_2$ localized in $J_0$, the adjoint action with
$\iota(w_{\gamma_1}w_{\gamma_2})$ takes $B_+(O_1)$ to
$B_+(\gamma\,O_1)$ if $O_1\subset O_0$. Again, it is sufficient to
verify the action on the charged intertwiners (\ref{chint1}) of
$B_+(O_1)$: 
\bea\label{diff}
\iota(w_{\gamma_1}w_{\gamma_2}) \cdot \psi_1 \cdot
\iota(w_{\gamma_1}w_{\gamma_2})^* = \iota\big((w_{\gamma_1}u\sigma(w_{\gamma_1}^*))(w_{\gamma_2}\bar u\bar\tau(w_{\gamma_2}^*))\big)\cdot \psi
\eea
where $w_{\gamma_1}u\sigma(w_{\gamma_1}^*)
\in\Hom(\sigma,\gamma_1\sigma_1\gamma_1\inv)$ and $w_{\gamma_2}\bar
u\bar\tau(w_{\gamma_2}^*)
\in\Hom(\bar\tau,\gamma_2\bar\tau_1\gamma_2\inv)$, and
$\gamma_1\sigma_1\gamma_1\inv$ is localized in $\gamma_1\,I_1$ and
$\gamma_2\bar\tau_1\gamma_2\inv$ is localized in
$\gamma_2\,J_1$. Hence (\ref{diff}) is a charged intertwiner of 
$B_+(\gamma O_1)$. This proves the claim. Then the diffeomorphism
covariance of $B_+$ and $B_{2D}$ follow because 
the diffeomorphisms localized in $O_0$ together with the M\"obius
group generate all diffeomorphisms.  \QED

\section{Cluster limit}\label{sec4}
\setcounter{equation}{0}

Let $b_1,\dots,b_n\in B_+(O)$ be BCFT observables localized within any
fixed double cone $O=I\times J\subset M_+$. We wish to consider the
behavior of a vacuum correlation  
\bea\label{b1bn} 
\big(\Omega\,,\, \beta_x(b_1\cdots b_n)\,\Omega\big),\eea
where $\beta_x=\beta^+_x\beta^-_{-x}$ is the one-parameter semigroup
of ``right shifts'' ($x>0$, away from the boundary), that take $I$ to $I+x$
and $J$ to $J-x$, represented as homomorphisms from $B_+(O)$
to $B_+(I+x\times J+x)$, see (\ref{beta}).

In Sect.~3 (with $O$ as the fixed reference
double cone) we have given the re-inter\-pretation of $b_i$ in the GNS
representation $\hat\pi$ of the state $\xi\circ\mu$ as observables of
the associated 2D CFT, with the 2D vacuum $\Omega_{2D}$ given by the
GNS vector. We shall show

\medskip 

{\bf Theorem 2:} {\sl Let each $b_i\in B_+(O)$ ($i=1,\dots,n)$ be of the
  form $\iota(a_1^{(i)} a_2^{(i)}) \cdot \psi^{(i)}$ with charged 
  intertwiners $\psi^{(i)}$ and $a^{(i)}_1\in A(I)$ and $a^{(i)}_2\in A(J)$. 
As $x$ goes to $+\infty$, the BCFT vacuum correlations (\ref{b1bn})
converge to the 2D vacuum correlations }
\bea\label{corr2d}
\big(\Omega_{2D}\,,\,\hat\pi(b_1\cdots
b_n)\,\Omega_{2D}\big) = \xi\circ\mu(b_1\cdots b_n). 
\eea

\medskip

{\it Proof:} We compute the limit and the 2D vacuum expectation value
separately. 

Using the decomposition of products $\psi_1\psi_2$ into
finite sums of operators of the form $\iota(T_1T_2)\cdot \psi$
\cite{LR2}, where $T_i$ are intertwiners between DHR 
endomorphisms of $A$, we see that the product $b_1\cdots b_n$ is a
finite sum of operators of the same form $\iota(a_1a_2)\cdot \psi$.

For the present purpose, it is more convenient to write the charged
intertwiners as $\psi= t\cdot\iota(\bar r)$ where $r\in
\Hom(\id,\tau\bar\tau)\subset A(J)$ and $t\in
\Hom(\alpha^+_\tau,\alpha^-_\sigma)\subset \Hom(\iota\tau,\iota\sigma)$
(Frobenius reciprocity). Then, because $a_2=\sigma(a_2)$, we get
$\iota(a_2)\cdot \psi = t\cdot \iota(\tau(a_2)\bar r)$. Hence, the product
$b_1\dots b_n$ is a finite sum of operators
of the form
\bea \label{ata}
\iota(a_1)\cdot t\cdot \iota(a_2).
\eea
Thus, the above vacuum correlation function is a finite sum
of expectation values
\bea F(x) = \big(\Omega\,,\,\beta_x\big(\iota(a_1)\cdot t\cdot
\iota(a_2)\big)\,\Omega\big) = \qquad\qquad\qquad\qquad \nonumber  
\\ = \big(\Omega\,,\,\iota(\alpha_x(a_1) z^\sigma(x))\cdot t\cdot 
\iota(z^\tau(-x)^* \alpha_{-x}(a_2))  \,\Omega\big) =  \\
= \big(\Omega\,,\,\alpha_x(a_1) z^\sigma(x)\cdot \eps(t)\cdot 
z^\tau(-x)^* \alpha_{-x}(a_2)\,\Omega\big). \nonumber \eea
Here, $\eps$ is the global conditional expectation $B\to A$, which
preserves the vacuum state \cite{LR1}. In particular, $\eps(t)\in
\Hom(\tau,\sigma)$. Therefore, the expression vanishes identically
unless $\sigma$ and $\tau$ belong to the same sector.

In the latter case, we express the cocycles as
$z^\rho(g)=U_0(g)U_\rho(g)^*$, and $\alpha_g={\mathrm {Ad}}_{U_0(g)}$, giving 
\bea
F(x) = \big(\Omega\,,\,a_1 U_\sigma(x)^* \cdot \eps(t) \cdot U_\tau(x)^*
a_2\,\Omega\big) =\big(\Omega\,,\,a_1 \cdot U_\sigma(-2x) \cdot \eps(t) a_2\,\Omega\big) ,
\eea
because the intertwiners between DHR endomorphisms also intertwine the 
representations of the M\"obius group \cite{FRS2}. 
By the spectrum condition, $F(x)$ has a bounded analytic
continuation to the lower complex halfplane. $U_\sigma(-z)$ weakly
converges in every direction $z=re^{i\varphi}$ ($-\pi < \varphi <0$,
$r\to\infty$) to the projection onto the zero eigenspace of the
generator, and the latter projection is nonzero only if $\sigma=\id$
is the vacuum representation; in this case $t=\eps(t)=1$. Thus, $F(z)$
converges in these directions to the vacuum expectation value 
\bea\label{limit}
\delta_{\sigma,0}\delta_{\tau,0}
\;(\Omega,a_1\Omega) \cdot(\Omega,a_2\Omega).
\eea

Next, we consider 
\bea
\overline{F(x)} = \big(\Omega\,,\,\beta_x\big(\iota(a_2^*)\cdot t^*\cdot
\iota(a_1^*)\big)\,\Omega\big).
\eea
Let $r_\sigma\in \Hom(\id,\bar\sigma\sigma)\subset A(I)$
and $r_\tau\in \Hom(\id,\bar\tau\tau)\subset A(J)$. Then we can write 
$t^* = \iota(r_\sigma^*)\cdot \bar t\cdot \iota(r_\tau)$, where 
$\bar t \in \Hom(\alpha^+_{\bar\tau},\alpha^-_{\bar\sigma})\subset 
\Hom(\iota\bar\tau,\iota\bar\sigma)$. Using the locality
properties of $a_1\in A(I)$, $a_2\in A(J)$, we can rewrite
\bea
\overline{F(x)} =
\big(\Omega\,,\,\beta_x\big(\iota(r_\sigma^*\bar\sigma(a_1^*))\cdot \bar t\cdot
\iota(\bar\tau(a_2^*)r_\tau)\big)\,\Omega\big). 
\eea
This expression can be computed in the same way as $F(x)$ before,
giving
\bea
\overline{F(x)} = \big(\Omega\,,\,
r_\sigma^*\bar\sigma(a_1^*))\cdot U_{\bar\sigma}(-2x) \cdot \eps(\bar t)
\bar\tau(a_2^*)r_\tau\,\Omega\big).
\eea
Thus $F(x)$ also has a bounded analytic continuation to the upper
complex halfplane, and converges to the same limit (\ref{limit}) also
in the directions $z=re^{i\varphi}$ ($0 < \varphi < \pi$, $r\to\infty$). 
From this, we may conclude the cluster limit  
\bea\label{cluster}
\lim_{x\to\infty}\big(\Omega\,,\,\beta_{x}\big(\iota(a_1)\cdot t\cdot
\iota(a_2)\big)\,\Omega\big) = \delta_{\sigma,0}\delta_{\tau,0}
\;(\Omega,a_1\Omega) \cdot(\Omega,a_2\Omega).
\eea

On the other hand, we now compute (\ref{corr2d}) and show that it
coincides with the factorizing cluster limit of
(\ref{b1bn}). For each contribution of the form
(\ref{ata}), we have
\bea
(\Omega_{2D}\,,\,\hat\pi\big(\iota(a_1)\cdot t\cdot
\iota(a_2)\big)\,,\,\Omega_{2D}) = \xi\circ\mu\big(\iota(a_1)\cdot t\cdot
\iota(a_2)\big) = \xi\big(a_1\cdot \mu(t)\cdot a_2\big). \qquad
\eea
But $\mu(t)\in A(I)\vee A(J)$ is an intertwiner in $\Hom(\sigma,\tau)$
which vanishes unless $\sigma=\id$ and $\tau=\id$ both belong to the
vacuum sector. In the latter case, $t=\mu(t)=1$. Thus,
\bea\label{cblock}
\langle\hat\Xi\vert \hat\pi\big(\iota(a_1)\cdot t\cdot
\iota(a_2)\big)\vert \hat\Xi\rangle = \delta_{\sigma,0}\delta_{\tau,0} \;
\xi(a_1a_2)  = \delta_{\sigma,0}\delta_{\tau,0} 
\;(\Omega,a_1\Omega) \cdot(\Omega,a_2\Omega). \qquad
\eea 
This coincides with the cluster limit (\ref{cluster}) ``far away from
the boundary''. \QED

\medskip 

Recall that $a_1$ and $a_2$ in (\ref{ata}) were obtained by
multiplying $b_1\cdots b_n$ and successively decomposing the products
of the charged intertwiners. Thus, the vacuum expectation values
$(\Omega,a_i\Omega)$ in (\ref{cblock}) are precisely the chiral
conformal blocks of the corresponding 2D correlation functions. 

A variant of the conformal cluster theorem \cite{FJ} should also give
a quantitative estimate for the rate of the convergence, depending on
the charges of the operators involved through the corresponding
spectrum of $L_0$.

\section{Conclusion}
We have studied the passage from a local conformal quantum field
theory defined on the halfspace $x>0$ of two-dimensional Minkowski
spacetime (boundary CFT, BCFT) to an associated local conformal
quantum field defined on the full Minkowski spacetime (2D CFT). There
are essentially two ways: the first is to consider BCFT vacuum
correlations of observables localized far away from the boundary. In
the limit of infinite distance, these correlation factorize into
chiral correlations (conformal blocks) of charged fields. We have
traced this effect back to the cluster property of the underlying
local chiral subtheory. 

The second method exploits the split property, i.e., the existence of
states of the underlying local chiral CFT in which correlations between
observables in two fixed intervals at a finite distance are suppressed. 
With the help of the split property one can algebraically identify a
fixed local algebra of the BCFT with a fixed local algebra of the 2D
CFT, and one can generate a unitary representation of the 2D M\"obius
group in the GNS Hilbert space of a suitable ``extended split state''
of this algebra. Its ground state is different from the BCFT
vacuum. Then, by acting with the 2D M\"obius group, one can obtain
{\em all} local algebras of the 2D CFT in the same Hilbert space. 

The converse question: can one consistently ``add'' a boundary in any
2D CFT (without affecting the algebraic structure away from the
boundary), is not addressed here. However, there arises a necessary 
condition from the discussion in App.~\ref{apphmi}: the 2D partition
function should be either modular invariant, or at least it should be
intermediate between the vacuum partition function and some modular
invariant partition function. We hope to return to this problem, and
find also a sufficient condition. 

\bigskip

\noindent
{\bf Acknowledgements:} KHR thanks the Dipartimento di Matematica of
the Universit\`a di Roma ``Tor Vergata'' for hospitality and financial
support, and M. Weiner and I. Runkel for discussions related to the subject.

\appendix

\section{Modular construction of $\Mb\times\Mb$ in the split state}
\label{appmod}
\setcounter{equation}{0}

In \cite{GLW} it was shown that a unitary representation of the
M\"obius group $\Mb$ is generated by the modular groups of a
``halfsided modular triple'', i.e., three von Neumann algebras $A_i$ 
($i=0,1,2$) with a joint cyclic and separating vector $\Psi$ such that
if $\sigma^i_t$ is the modular group for $(A_i,\Psi)$, then
$\sigma^i_t(A_{i+1})\subset A_{i+1}$ for $t\leq 0$. (Here, $i+1$ is
understood mod 3.) Specifically, when $I$ is an open interval
and $I_1,I_2$ are the subintervals obtained by removing an interior
point from $I$, the three algebras $A_1=A(I_1)$, $A_2=A(I_2)$,
$A_3=A(I)'$ in a local chiral CFT together with the vacuum vector
$\Omega$ define a halfsided modular triple. This means that the entire
local net can be recovered from these data. 

We want to show here, how this construction can be applied to
construct a unitary representation of the 2D M\"obius group
$\Mb\times\Mb$ from six suitable algebras in the split state $\Xi$
associated with a pair of intervals $I$ and $J$, see (\ref{splitstate}).

\medskip 

Let $I_1,I_2$ arise from $I$ be removing a point, and similarly
$J_1,J_2$. Tensoring by 1, the two halfsided modular triples  
\bea
\big(\,A(I)'\otimes 1\,,\,A(I_1)\otimes 1\,,\, A(I_2)\otimes 1\, \big) 
\nonumber \\ 
\big(\, 1 \otimes A(J)'\,,\, 1 \otimes A(J_1)\,,\,1 \otimes A(J_2)\,\big)
\eea
in the state $\Omega\otimes\Omega$ generate $U_0\otimes U_0$. Under the
split isomorphism, these triples turn into 
\bea
\big(A(I)'\cap N,A(I_1),A(I_2)\big),\qquad
\big(A(J)'\cap N',A(J_1),A(J_2)\big)
\eea
in the split state $\Xi$, where $N$ is the canonical intermediate
type $I$ factor between $A(I)$ and $A(J)'$. $\Xi$ is cyclic and
separating for these algebras in the subspaces $\overline{N\Xi}$ and
$\overline{N'\Xi}$, respectively. The latter halfsided modular triples
thus generate the two commuting representations $U_+,U_-$ of $\Mb$
directly in $\HH_0$.   

\section{Charged intertwiners in BCFT}
\label{appeig}
\setcounter{equation}{0}

The charged intertwiners $\psi$ for a given chiral extension $A\subset
B$, that together with $A_+(O)$ generate $B_+(O)$, are
elements of the finite-dimensional spaces
$\Hom(\iota,\iota\sigma\bar\tau)\cap B(K)'$. In \cite[Eq.\ (5.12)]{LR2} a
linear condition on $\varphi=\bar\iota(\psi)\in
\Hom(\theta,\theta\sigma\bar\tau)$ was given which guarantees
that $\varphi$ commutes with $\bar\iota(B(K))$. Here $\bar\iota:B\to
A$ is a homomorphism conjugate to the injection $\iota:A\to B$, such
that $\gamma=\iota\bar\iota$ on $B(K)$ is a canonical endomorphism for
$A(K)\subset B(K)$ and $\theta=\bar\iota\iota$ is the dual canonical
endomorphism, which is a DHR endomorphism of $A$ localized in $K$
\cite{LR1}.  

Unfortunately, the condition displayed in \cite{LR2} does not
take into account that $\varphi$ belongs to $\bar\iota(B(L))$ (i.e.,
is in the range of $\bar\iota$). We want to reformulate this condition
so that it is equivalent to $\psi$ belonging to $B_+(O)=B(K)'\cap B(L)$. 

\medskip 

We first notice that every element of $B(K)$ is of the form 
$\psi= \iota(y)\, v$ where $v\in \Hom(\id_B,\gamma)\subset B(K)$ is
the canonical isometry intertwining $\gamma$. Then $\psi \in
\Hom(\iota,\iota\sigma\bar\tau)$ if and only if $y\in
\Hom(\theta,\sigma\bar\tau)\subset A(L)$. This already secures that
$\psi\in B(L)$, and since $\theta$ is localized in $K$, $\psi$ commutes with
$\iota(A(K))$. Hence it commutes with $B(K)$ iff it also commutes with
$v\in\Hom(\id_B,\gamma)$. This is equivalent to the relation 
\bea\label{comm} 
y\,x \stackrel!= \theta(y)\,x \equiv
\sigma(\eps_{\theta,\bar\tau})\,\eps_{\sigma,\theta}^*  \;
\theta(y)\,x \eea 
where $x=\bar\iota(v) \in \Hom(\theta,\theta^2)$. The statistics
operators $\eps$ are trivial \cite{FRS1} due to the localizations of
$\sigma$ in $I$, $\bar\tau$ in $J$, and $\theta$ in $K$, but we have
displayed them in order to make the condition covariant under unitary
deformations of $\bar\iota$ and $v\in\Hom(\id_B,\iota\bar\iota)$,
possibly changing the localization of $\theta$ and leading to
nontrivial statistics operators.

The condition (\ref{comm}) can be equivalently written as the
eigenvalue equation   
\bea\label{eigen}
\Pi(y):= \lambda^{\frac12} \cdot \big(1_{\sigma}\times r^* \times
1_{\bar\tau}\big)\circ
\big(\eps_{\sigma,\theta}^*\times\eps_{\theta,\bar\tau}^*\big)\circ 
\big(1_\theta\times y\times 1_\theta\big)\circ x_2 \stackrel!= y.
\eea
Here $r= x\circ w\in \Hom(\id_A,\theta^2)$ where $w\in
\Hom(\id_A,\theta)\subset A(K)$ is the dual canonical isometry (such
that $(\gamma,v,\iota(w))$ form a Q-system); $x_2=(1_\theta\times
x)\circ x = (x\times 1_\theta)\circ x \in \Hom(\theta,\theta^3)$, and
$\lambda\geq 1$ is the index $[B:A]$. $\circ$ and $\times$ are the
concatenation and  the monoidal product
in the tensor category of DHR endomorphisms of $A$. The map $\Pi$ defined by
(\ref{eigen}) is a linear map $\Pi:\Hom(\theta,\sigma\bar\tau)\to
\Hom(\theta,\sigma\bar\tau)$. (\ref{eigen}) obviously follows from 
(\ref{comm}) by left multiplication with $\sigma(r^*)$ and right
multiplication with $x$. To see that (\ref{eigen}) implies
(\ref{comm}), one may insert (\ref{eigen}) into both sides of
(\ref{comm}) and repeatedly use the relations of the dual Q-system 
$(\theta,w,x)$ to get equality. 

We thank I. Runkel who has pointed out to us that $\Pi$ is in fact a 
projection. Hence the charged intertwiners $\psi$ are precisely given
by $\iota(y)\cdot v$ where $y$ is in the range of $\Pi$. 
The multiplicities $Z_{\sigma,\tau}$ in (\ref{dim}) equal the dimension of
the range of these projections (for each pair $\sigma,\tau$). 

\section{Haag duality and modular invariance}
\label{apphmi}
\setcounter{equation}{0}

If $A$ is completely rational, the C* tensor category defined by
its DHR super\-selection sectors is modular \cite{KLM}, i.e., the
unitary $S$ and $T$ matrices defined by the statistics \cite{FRS2}
generate a representation of the group $SL(2,\ZZ)$. By the Verlinde
formula \cite{V}, these matrices also describe the modular transformation
behavior of chiral partition functions (``characters''). 

By \cite{BEK}, the matrix $Z$ given by (\ref{dim}) is a modular 
invariant (it commutes with $S$ and $T$), hence the partition function
of the 2D CFT $B_{2D}$ on $\HH_{2D}$ is invariant under modular transformations. 
We want to point out an interesting relation of this fact to Haag
duality of the associated BCFT. 

\medskip

As mentioned before, every BCFT defined by (\ref{relcomm}) is
automatically Haag dual, and any non Haag dual BCFT $\widetilde B_+$
with the same chiral observables is intermediate between $A_+$ and
$B_+$ \cite{LR2}. Therefore, the charged intertwiners $\psi\in 
\widetilde B_+$ constitute linear subspaces of the spaces of 
charged intertwiners in $B_+$. Let the dimensions of these spaces be
$\widetilde Z_{\sigma,\tau} \leq Z_{\sigma,\tau}$, and at least one of them $<
Z_{\sigma,\tau}$ (i.e., $\widetilde B_+$ is strictly contained in $B_+$). Then
the matrix $\widetilde Z$ cannot be a modular invariant by the
following simple argument: consider the $00$ component of
$S^*\widetilde Z S$. Because each $S_{i0}$ is positive, 
\bea
(S^*\widetilde Z S)_{00} = \sum_{ij} S_{0i}S_{0j} \widetilde Z_{ij}
\eea 
is strictly smaller than $(S^*ZS)_{00} = Z_{00} = 1$. If $\widetilde
Z$ were modular invariant, we would conclude $\widetilde Z_{00} < 1$,
which is impossible.  

We notice that the construction of a 2D CFT associated to a BCFT
described in Sect.~3 takes an intermediate BCFT
$A_+\subset \widetilde B_+\subset B_+$ to an intermediate 2D CFT
$A_{2D} \subset \widetilde B_{2D} \subset B_{2D}$. Its
Hilbert space is of the form (\ref{h2d}) with $Z$ replaced by
$\widetilde Z$. Hence, we conclude that the partition function of the
associated 2D CFT is modular invariant if and only if the BCFT is Haag dual.

\end{document}